\documentclass{PoS}

\usepackage{graphicx}
\usepackage{amsmath}

\newcommand{\bee}{\begin{equation}}
\newcommand{\ee}{\end{equation}}
\newcommand{\beea}{\begin{eqnarray}}
\newcommand{\eea}{\end{eqnarray}}

\def\tr{{\rm tr}}

\newcommand{\bea}{\begin{eqnarray}}
\newcommand{\ba}{\begin{array}}
\newcommand{\ea}{\end{array}}

\title{Algorithms for dynamical overlap fermions\footnote{Preprint number: DESY 06-175}}

\ShortTitle{Algorithms for dynamical overlap fermions}

\author{\speaker{Stefan Schaefer}\\
 NIC, DESY \\
 Platanenallee 6\\
 15738 Zeuthen\\
 Germany\\
 E-mail: \email{Stefan.Schaefer@desy.de}
}

\abstract{
An overview of the current status of algorithmic approaches to dynamical
overlap fermions is given. In particular the issue of  changing the 
topological sector is discussed.
}

\FullConference{XXIVth International Symposium on Lattice Field Theory\\
                 Tucson, Arizona}

\begin{document}
\section{Introduction}
Symmetries play a crucial role in our understanding of strong interactions physics.
Of particular interest is the $SU(N_f)\times SU(N_f)$ chiral symmetry of the 
massless QCD Lagrangian. Its spontaneous and explicit symmetry breaking pattern
 is the starting point for a large part of low-energy phenomenology.
Furthermore, an index theorem holds for the continuum Dirac operator which
 is believed to have significant consequences for low-energy physics too.
This theorem states that the index $\nu$ of the  Dirac operator  is equal
to the topological charge $Q_{\rm top}$ of the gauge field 
\bee
\nu = n_L-n_R=\frac{1}{2}{\rm tr} \, \gamma_5 D(0) = Q_{\rm top} 
\label{eq:index}
\ee
with $n_L$ and $n_R$ respectively the number of left and right handed zero-modes
of the massless Dirac operator $D(0)$.
Since these properties are  important in the continuum, it
is only natural to respect chiral symmetry when formulating QCD on the lattice.

The correct 
form of chiral symmetry at finite lattice spacing~\cite{Luscher:1998pq} is 
 the Ginsparg--Wilson (GW) relation~\cite{Ginsparg:1981bj} 
\bee
D(0) \gamma_5 + \gamma_5 D(0)= \frac{a}{R_0}D(0) \gamma_5  D(0) 
\label{eq:GW}
\ee
with $a$ the lattice spacing and $R_0$ a  parameter. 
For operators which fulfill the GW equation the index theorem~\cite{Hasenfratz:1998ri} 
works at finite lattice spacing just like in the continuum.
Moreover, the theory is automatically ${\cal O}(a)$ improved,
additive mass renormalization is absent and renormalization patterns can
be greatly simplified~\cite{Hasenfratz:1998jp}. The fermion determinant
is strictly positive if the quark mass is greater than zero.
Prominent solutions of the Ginsparg--Wilson equation include
domain wall fermions at large extent of the 
fifth dimension~\cite{Kaplan:1992bt, Shamir:1993zy},  the overlap operator~\cite{Neuberger:1997fp}
and the  fixed point Dirac operator~\cite{Hasenfratz:1993sp,Hasenfratz:2000xz}.
Unfortunately, the application of these operators is significantly
more expensive than standard  operators like the Wilson or staggered
Dirac operator. This prevented so far a more wide-spread use.

In recent years  simulations of QCD on large lattices and at
small quark masses have become possible. This is not only due
to increased computer resources but rather because of algorithmic advances,
see the reviews given at this conference\cite{ClarkPlenary,GiustiPlenary}.
 These improvements apply to the simulation of dynamical
chiral fermions too. In one respect, however, chiral fermions are different.
Because of the index theorem Eq.~\ref{eq:index}, any  chiral Dirac operator 
changes discontinuously
where its index changes. This has to be accounted for in the equations of motion
which are at the heart of most current algorithms. 
The different approaches how to deal with the change of topological sector
algorithmically are the main subject of this write-up.

This proceedings contribution is about dynamical simulations using Neuberger's 
overlap Dirac operator. However, also domain wall fermions at finite time extent, 
the parameterized  fixed point operator~\cite{Hasenfratz:2005tt} 
and chirally improved fermions~\cite{Lang:2005jz},
which all implement chiral symmetry to a high degree,
have been used in dynamical simulations. Their approximate symmetry
leads to an approximation of the discontinuity encountered for actions with
exact chiral symmetry. As we are going to discuss, this can cause large
forces in the molecular dynamics evolution while changing topological sector.
The issue of changing topology has therefore received considerable attention
in all those simulations too.

This paper is organized as follows. First we review the definition of the 
overlap operator and the fermionic action with a focus on the discontinuity
where the index  changes.
Three methods to deal with the discontinuity in the context of 
Hybrid Monte Carlo are discussed: modified molecular dynamics evolution in Sec.~\ref{sec:FKS},
approximation of the discontinuity in Sec.~\ref{sec:approx} and  
topology conserving actions in Sec.~\ref{sec:constopo} . The write-up closes with a summary.

\section{The action\label{sec:action}}
Let us start by reviewing the action whose simulation is the subject of this paper.
The overlap Dirac operator~\cite{Neuberger:1997fp} is given by
\bee
D_{\rm ov}(m) = (R_0-\frac{m}{2}) \big[ 1+ \gamma_5 \epsilon ( h(-R_0)) \big] +m
\label{eq:dov}
\ee
with $m$ the bare quark mass.
It is constructed from a Hermitian lattice Dirac operator $h=\gamma_5 d$ which
is typically the standard Wilson operator. It is taken at the negative mass $-R_0$
 at the cut-off scale. $R_0$ can be tuned to achieve optimal locality
of the resulting overlap operator~\cite{Hernandez:1998et}. In the following $h(-R_0)$ is referred to 
as the kernel operator.
$\epsilon(h)$ is the matrix sign function. For Hermitian, non-singular 
matrices it can be defined by
\bee
\epsilon(h)=\frac{h}{\sqrt{h^2}}=\sum_i {\rm sign} (\lambda_i) \psi^{}_i \psi^\dagger_i
\label{eq:sign}
\ee
with the sum running over all eigenvalues $\lambda_i$ of the kernel operator $h$ and
$\psi_i$ the associated eigenmodes. The numerical construction of $\epsilon(h)$ usually
uses a combination of the two definitions in Eq.~\ref{eq:sign}. The lowest modes of $h(-R_0)$
are computed explicitly and the spectral representation can be used. For the 
rest of the spectrum a polynomial or rational approximation to $1/\sqrt{h^2}$ implements
the sign function to very high accuracy.

For simulations of two flavors one is also interested in $H^2=D^\dagger D$, with $H=\gamma_5 D$.
(Here and in the following we denote the overlap Dirac operator by capital letters $D$ and $H$
whereas the Hermitian  kernel operator is denoted by small $h$.)
Because of the Ginsparg--Wilson equation and $\gamma_5$-Hermiticity, $H^2$ commutes
with $\gamma_5$ and is therefore block-diagonal in the chiralities with
\bee
H^2_\sigma=P_\sigma H^2 P_\sigma= 
2(R_0^2-\frac{m^2}{4}) P_\sigma \big[ 1+\sigma \epsilon ( h(-R_0)) \big]P_\sigma +m^2 P_\sigma
\ee
where $P_\sigma=\frac{1}{2}(1+\sigma \gamma_5)$ projects to  chirality $\sigma=\pm 1$.

Obviously, the overlap operator is discontinuous where one of the eigenvalues $\lambda_0$
of the kernel operator changes sign. This is precisely where the topological charge
as defined by the index of the Dirac operator changes. Evaluating the lattice version of Eq.~\ref{eq:index} for 
the overlap operator~\cite{Hasenfratz:1998ri} gives
\bee
\nu=\frac{1}{2R_0} {\rm Tr} \gamma_5 D(m=0) = \frac{1}{2} {\rm Tr} \epsilon( h(-R_0) ) = 
\frac{1}{2}\sum_i {\rm sign} \lambda_i 
\ee
with the sum running over all eigenvalues $\lambda_i$ of the kernel operator. 
The direction of the sign change determines whether the topological
charge decreases or increases by one unit. 

On the surfaces which separate the topological sectors, 
the spectrum of the overlap operator changes: One zero-mode 
is created or vanishes and also the rest of the spectrum moves.
Interestingly, the ratio of the fermion determinants on the two sides
of the step can be computed
without the explicit computation of the determinant itself:
Because the eigenmodes of the kernel are continuous functions
of the gauge fields, the change of the sign of one of the eigenvalues
amounts to the following change in the overlap operator
\bee
\tilde D(m) = D(m) \, - \, (2R_0-m)\,  {\rm sign} (\lambda_0) \,  \gamma_5  \psi_0 \psi_0^\dagger
\ee
with ${\rm sign} \lambda_0$ the sign of the eigenvalue before the 
crossing and $\psi_0$ the associated eigenfunction.
This  translates to a discontinuity of the fermion determinant~\cite{DeGrand:2005vb}
\bee
\det \tilde D(m) =\left[ 1 -  (2R_0-m)\,  {\rm sign} (\lambda_0) \, 
\big(\psi_0^\dagger \frac{1}{\gamma_5 D(m)} \psi_0\big) \right]   \det D(m) 
\label{eq:deltadet}
\ee
which can be computed by just one inversion of the overlap operator on the crossing mode.
However, most of the time we do not deal with the determinant directly but introduce it
into the functional integral by pseudo-fermion fields
\bee
\det H^2 \propto \int {\rm d} \phi \, {\rm d} \phi^\dagger \  e^{-\phi^\dagger H^{-2} \phi} =
\int {\rm d} \phi \, {\rm d} \phi^\dagger \  e^{-S_{f}[U,\phi]}  \label{eq:PF}
\ee
with $S_f[\phi, U]$ the effective fermion action. 
The discontinuity in $S_f[\phi,U]$ can be computed using the Sherman--Morrison formula
\bee
\Delta S_{f}=\Delta \sum_\sigma \phi^\dagger \frac{1}{H_\sigma^2} \phi 
=\sum_{\sigma=\pm 1} \sigma{\rm sign}(\lambda_0)
\frac{(4R_0^2-m^2)}
{1- \sigma {\rm sign}(\lambda_0) (4R_0^2-m^2) \psi_0 H^{-2}_\sigma(m)\psi_0}
|\phi \frac{1}{H_\sigma(m)^2}\psi_0|^2 
\label{eq:sherman}
\ee
which again can be evaluated by inverting the (squared) overlap operator on the crossing mode of
the kernel operator.

In summary, the overlap operator is constructed via the matrix sign function
of a doubler free lattice Dirac operator. This makes its application roughly 20--200 times
more expensive than a standard Wilson operator.\footnote{A recent comparison\cite{Chiarappa:2006hz} finds
the cost of computing the propagator with the overlap operator to be 30--120 times larger than with
twisted mass fermions.} The fermion determinant and the effective 
action for the overlap operator are non-zero and  finite everywhere. However, it has discontinuities
where the topological charge (the index) changes. They have their origin in one eigenvalue
of the kernel operator changing sign. The height of these steps can 
be computed by one inversion of $D^\dagger D$ on the crossing mode.

\section{Hybrid Monte Carlo}
The most popular exact algorithm used in dynamical fermion simulations is Hybrid
Monte Carlo \cite{Duane:1987de}. It is a combination of Molecular 
Dynamics (MD)~\cite{Gottlieb:1987mq} with a Metropolis accept/reject 
step that makes it exact.
Like in all MD based simulations conjugate momenta $\pi$ are introduced which drive
the evolution of the generalized coordinates, the gauge fields in our case.
The Hamiltonian $H[U,\pi,\phi]$ from which the equations of motion are derived is composed
of the kinetic term for the momenta and the potential given by the 
action of the theory which we actually want to simulate
\bee
H[U,\pi,\phi] = \frac{\pi^2}{2}+S_f[U,\phi]+S_g[U] \ .
\ee
$S_f[U,\phi]$ is  the effective fermion action at fixed pseudo-fermion fields and $S_g[U]$
the gauge action.

The elementary update step in the HMC algorithm is called a trajectory. At the beginning, a heat bath for the 
pseudo-fermion field and the momenta is performed. Then the equations of motion,
which derive from ${\rm d} H / {\rm d} \tau=0$ are
solved numerically for some (fictitious) time $\tau$.
 The simplest integrator is the leap frog which repeats
\bee
\big[U \longrightarrow T_U(\delta \tau /2 ) U \ \ , \ \
\pi \longrightarrow T_p(\delta \tau  ) \pi  \ \ , \ \
U \longrightarrow T_U(\delta \tau /2 ) U  \big ]
\ee
$\tau/\delta \tau$ times; $T_U$ and $T_p$ are the gauge field and momentum updates. 
 At the end of the trajectory, an
accept/reject step is performed which corrects for errors introduced by the numerical
integration. 
The new gauge configuration is accepted with probability
$P_{\rm acc}={\rm min} \{\exp (-\Delta H [U,\pi,\phi]) , 1 \}$.

The leap-frog integrator provides us with an approximate solution to the equations of 
motion with an error that vanishes with $(\delta \tau)^2$ as $\delta \tau \to 0$. 
Decreasing the step-size can therefore bring the acceptance rate arbitrarily close to 1.
This is not the case for the overlap action because the leap-frog (almost) never
hits the discontinuity and its effect is therefore not taken into account
in the numerical integration.
Without a special treatment of the step in the action, the acceptance rate is therefore smaller
than 1; typically far too small for any practical purpose.
How to deal with this discontinuity is the subject of the rest of this write-up.
In Sec.~\ref{sec:FKS}, we discuss the modification of the MD evolution proposed by 
Fodor, Katz and Szabo (FKS)~\cite{Fodor:2003bh,Fodor:2004wx}. The subject of  Sec.~\ref{sec:approx} are algorithms
which approximate the step in the action and thereby make the evolution accessible to standard
methods. This approximation has the to be corrected for to get the exact overlap action.
Topology conserving actions, which solve the problem by avoiding the discontinuity completely,
are described in Sec.~\ref{sec:constopo}.
First, however, we discuss various possibilities to introduce the fermion determinant into
the simulation.

\subsection{Multiple pseudo fermions}
In recent years, the quality of the estimator used to introduce the fermion determinant 
into the functional integral has turned out to be of crucial importance to the performance of
standard algorithms.  The pseudo-fermion field in Eq.~\ref{eq:PF}
is such a stochastic estimator for the change in the fermion determinant over an update:
At the starting configuration a pseudo-fermion heat-bath
is performed, $\phi=D(m) R$ with $R$ a Gaussian random field. The change in the determinant
can be computed by the average over the random field
\bee
\frac{\det \tilde H^2}{\det H^2}=\langle e^{-\Delta S_f[\phi,U]} \rangle \ .
\ee
Using just one pseudo-fermion field for this estimate is known to be very noisy~\cite{Hasenfratz:2002ym}
which in the context of Hybrid Monte Carlo can cause large forces.

A successful way of improvement is to replace the original fermion matrix by a product of matrices
\bee
\det Q =\det D^\dagger D= \det M_1 \det M_2 \cdots \det M_N \, .
\ee
Each of these determinants is then introduced by a pseudo-fermion field.
If the matrices $M_i$ are suitably chosen, 
it can be easier to estimate the (change in) their determinant stochastically and the
fluctuations are greatly reduced.

The initial idea by Hasenbusch was to use $N=2$, $M_1=Q(m)/Q(M)$ and $M_2=Q(M)$ with $Q(M)$ the
fermion matrix at a larger quark mass~\cite{Hasenbusch:2001ne,Hasenbusch:2002ai}. Another 
choice is to use the $N$-th root of $Q$ for all $M_i$~\cite{Hasenbusch:1998yb,Clark:2006fx}. 
So far neither of these two methods has proved superior to the other.
The former, however, is easier to implement. 
As we will see in the following, the improved estimator not only reduces the fermionic
forces but it 
can also drastically reduce the auto-correlation time of the topological charge~\cite{DeGrand:2004nq}.

\subsection{Modified evolution\label{sec:FKS}}
Most of the experience with dynamical overlap fermions is based on the
algorithm proposed by Fodor, Katz and Szabo (FKS)~\cite{Fodor:2003bh},
which is used---sometimes in variants---in
Refs.~\cite{DeGrand:2004nq,Cundy:2004xf,Cundy:2005pi,DeGrand:2005vb,Schaefer:2005qg,Cundy:2005mn,DeGrand:2006ws,DeGrand:2006uy,DeGrand:2006nv}.
So far the simulations have been limited to small volume, lattices up to $10^4$
sites albeit at a coarse lattice spacing.

The FKS algorithm is the standard Hybrid Monte Carlo  unless a discontinuity is encountered. 
This can only happen---an eigenvalue of the kernel operator $h(-R_0)$ changes sign---during 
the gauge field update $T_U(\delta\tau/2)$. The  
corresponding update step is the split in two: one where the trajectory is advanced right to the point
where the index changes by $T_U(\delta \tau_1)$, i.e. 
where the eigenvalue of the kernel operator gets  zero.
There a discrete update of the momenta is performed. It depends on whether
there is enough momentum  $\pi_\perp$ perpendicular to the surface which separates the 
two topological sectors to get across the step in the action $\Delta S$. In this case,
the momentum is 
reduced such that energy is (approximately) conserved and the topological sector changes. 
If there is not enough
momentum, $\pi_\perp$ is reversed, the trajectory is reflected off the surface and 
stays in the same topological sector. To be precise, the momentum component $\pi_\perp$ 
is changed to $\pi_\perp'$ in the following way
\bee
\pi'_\perp=
\begin{cases}
-\pi_\perp & \text{if} \ \  2\Delta S > |\pi_\perp|^2 \\
\pi_\perp \sqrt{1-\frac{2\Delta S}{|\pi_\perp|^2}}& \text{if} \ \  2\Delta S < |\pi_\perp|^2 
\end{cases}
\ee
After that, the gauge fields are advanced to the 
end of the update step by  $T_U(\delta \tau/2-\delta \tau_1)$.
This update is obviously reversible and was proved to be area conserving too~\cite{Fodor:2003bh}.
Combined with the standard Metropolis step at the end of the trajectory,
the resulting algorithm is exact. Improvements of the original update are discussed in~\cite{Cundy:2005mn}.

Here are a few more details:
The component of the momentum perpendicular  to the  $\lambda=0$ surface 
is computed using the Feynman--Hellmann theorem
$\delta \lambda_0/\delta U = \psi^\dagger_0 (\delta h/\delta U) \psi_0$
with $\psi_0$ the crossing eigenmode of the kernel operator $h$ and $\lambda_0$ its eigenvalue.
Thus 
\bee
A=\left[ (\psi^\dagger_0 \frac{{\delta}h}{{\delta} U} ) \otimes U \psi_0 \right]_{TA}
\ee
is perpendicular to the 
surface separating the topological sectors. TA denotes the traceless, anti-hermitian component of the vector.
With $N=A/|A|$ the normalized vector perpendicular to the surface separating the two topological
sectors, the momentum perpendicular to the surface is simply $\pi_\perp=N (\pi\cdot N)$.
The height of the step $\Delta S$ is computed using Eq.~\ref{eq:sherman} which requires
one inversion of the overlap operator on the crossing mode, independently of how many pseudo-fermion
fields are used.
\begin{figure}
\begin{center}
\includegraphics[width=0.28\textwidth,angle=-90]{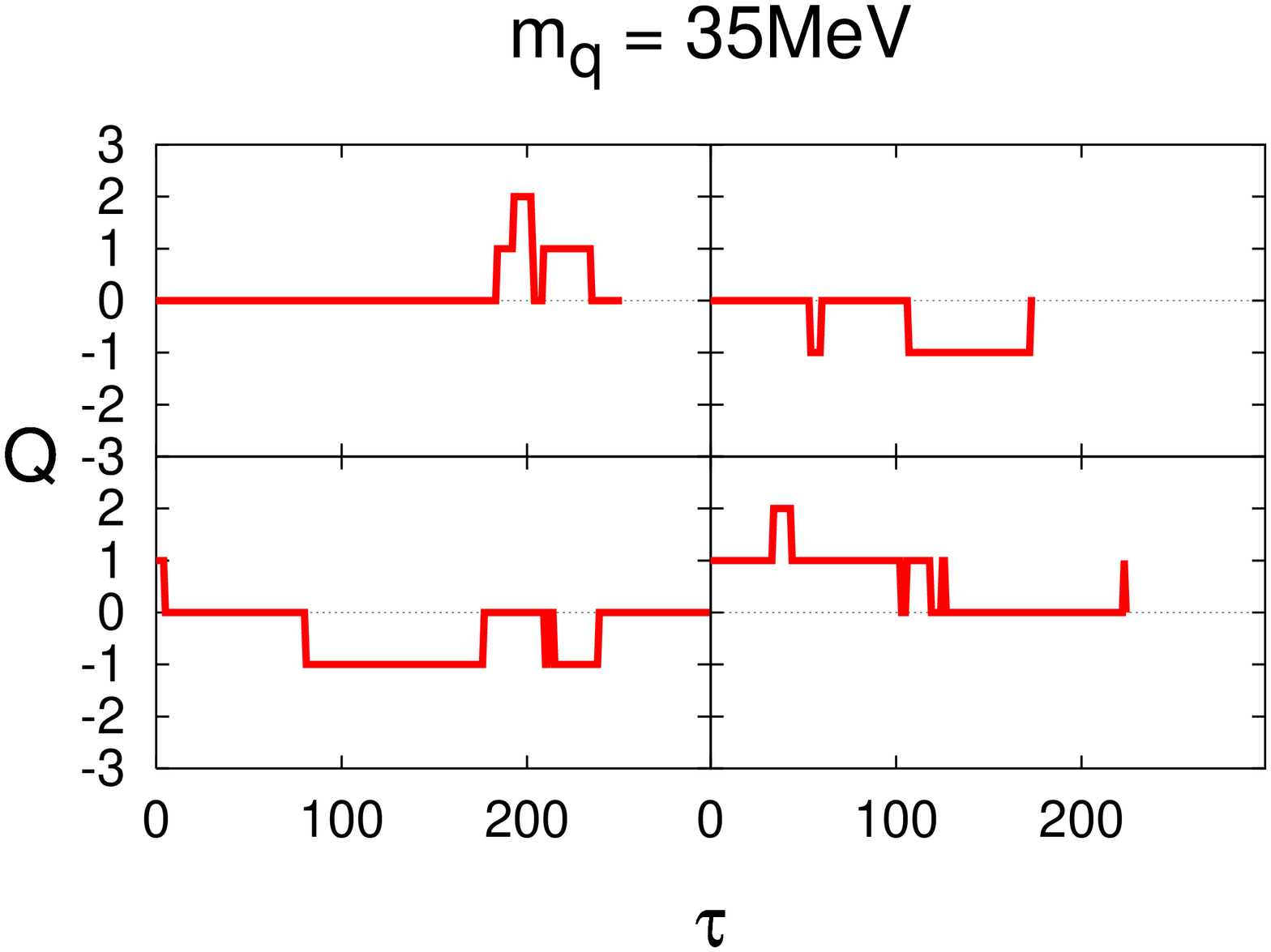}
\includegraphics[width=0.28\textwidth,angle=-90]{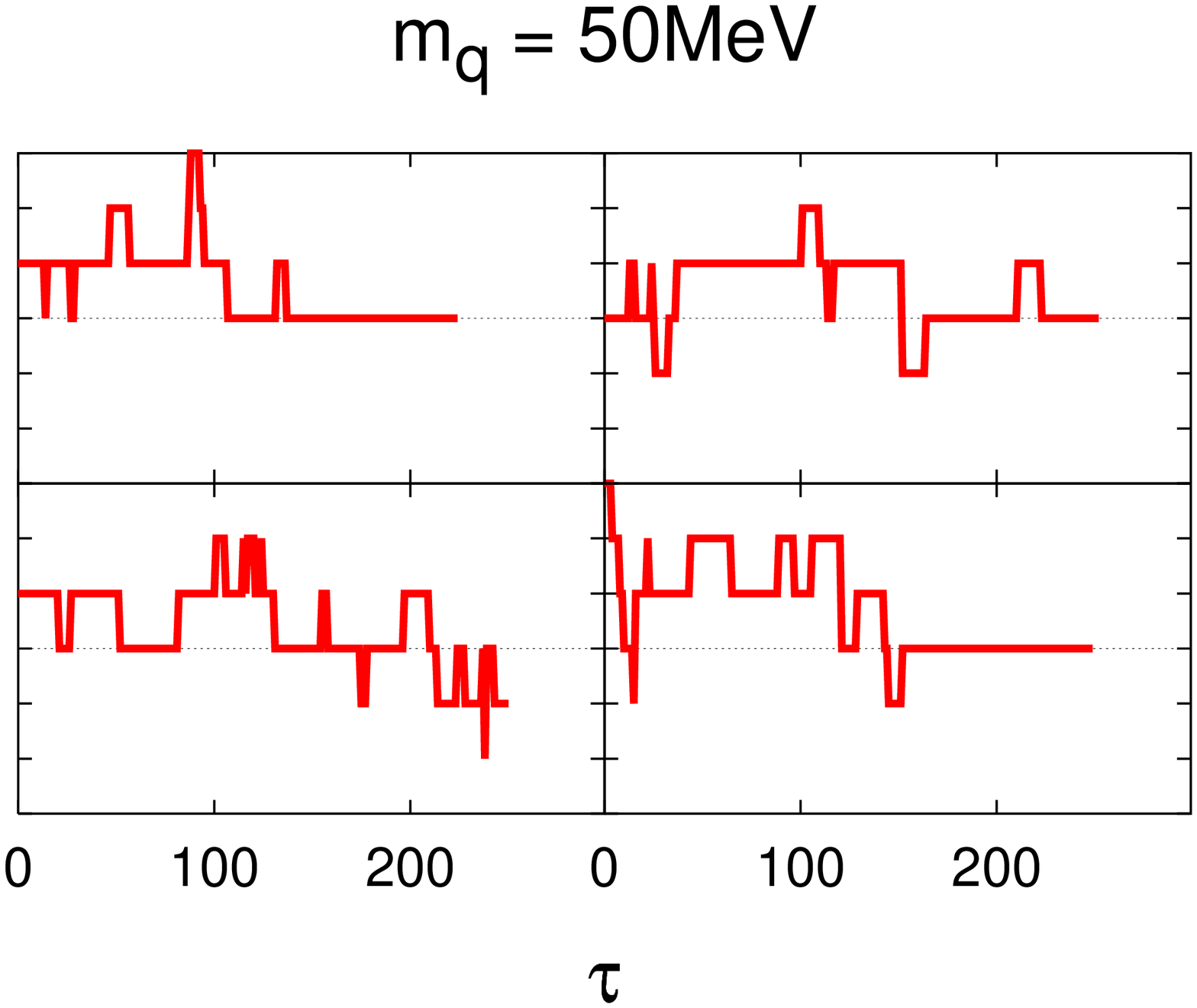}
\includegraphics[width=0.28\textwidth,angle=-90]{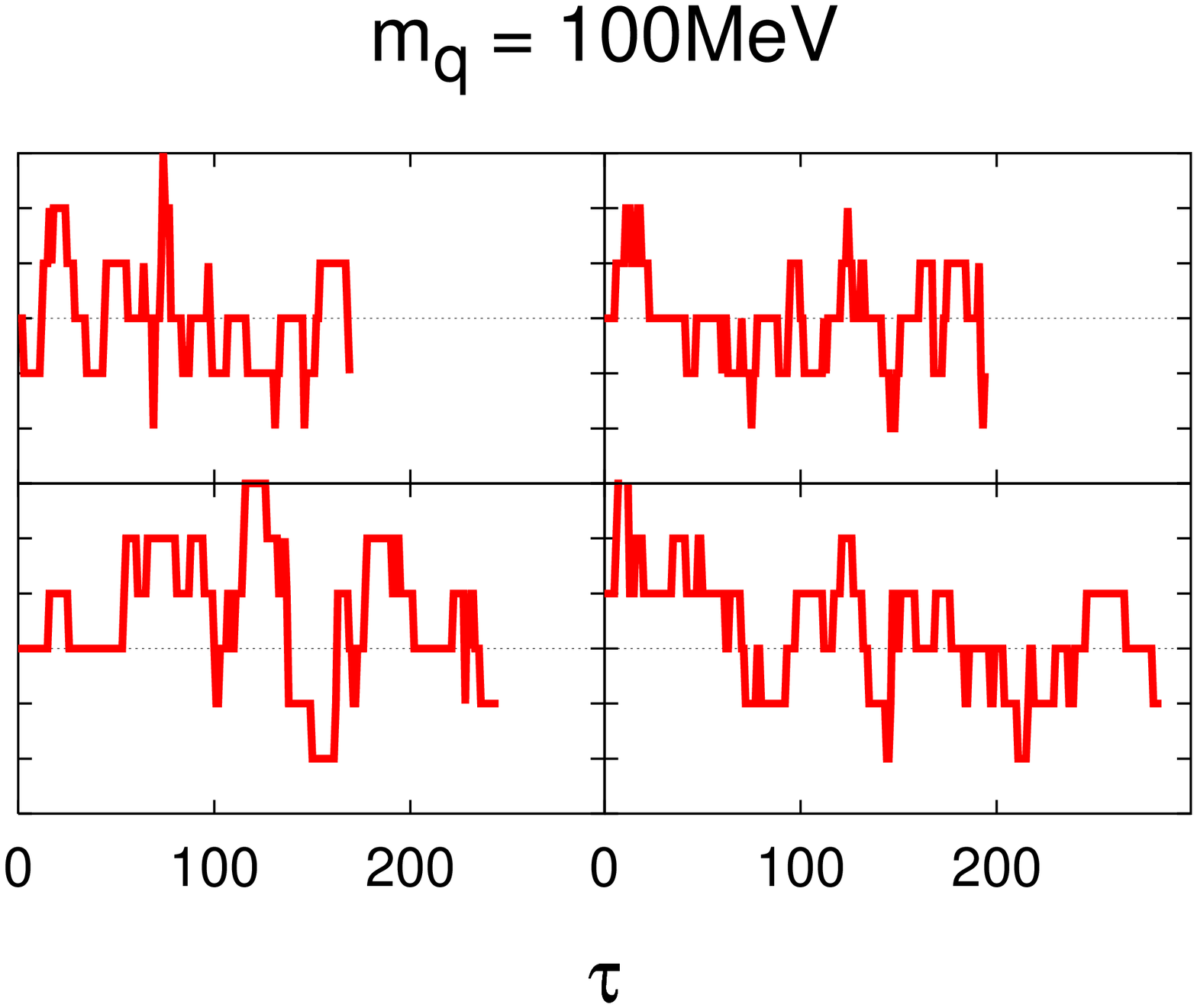}
\end{center}
\caption{The history of the topological charge as a function of simulation time. The data
is from simulations on a $8^4$ lattice at lattice 
spacing $a\approx 0.15$~fm \cite{DeGrand:2005vb}.\label{fig:qq}}
\end{figure}

\begin{figure}
\begin{center}
\includegraphics[width=0.3\textwidth,angle=-90]{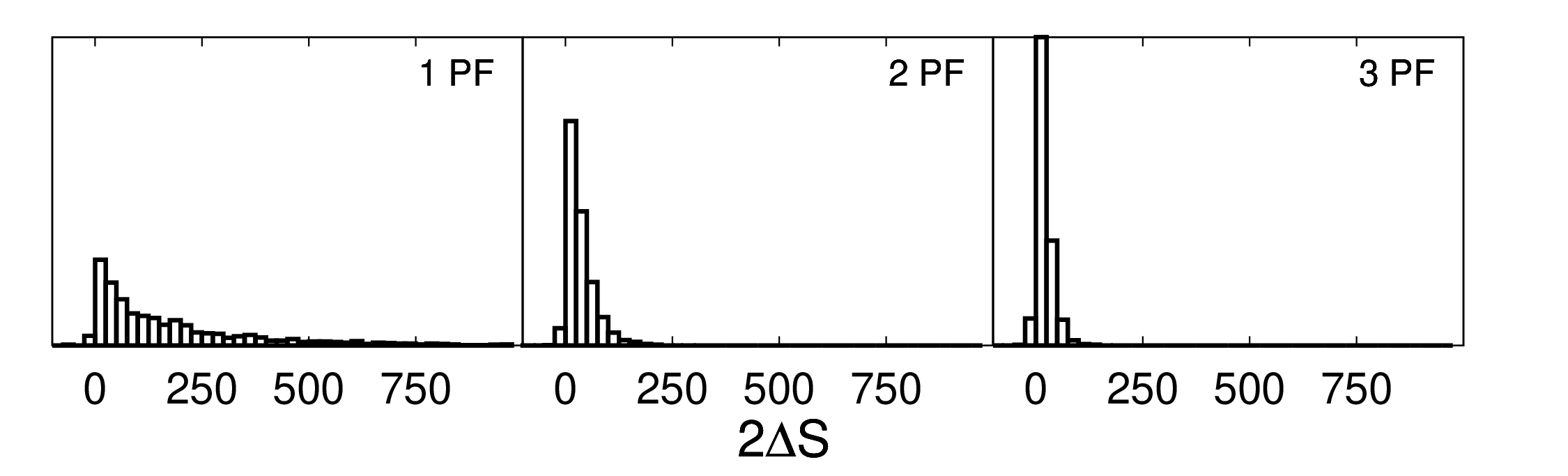}
\includegraphics[width=0.3\textwidth,angle=-90]{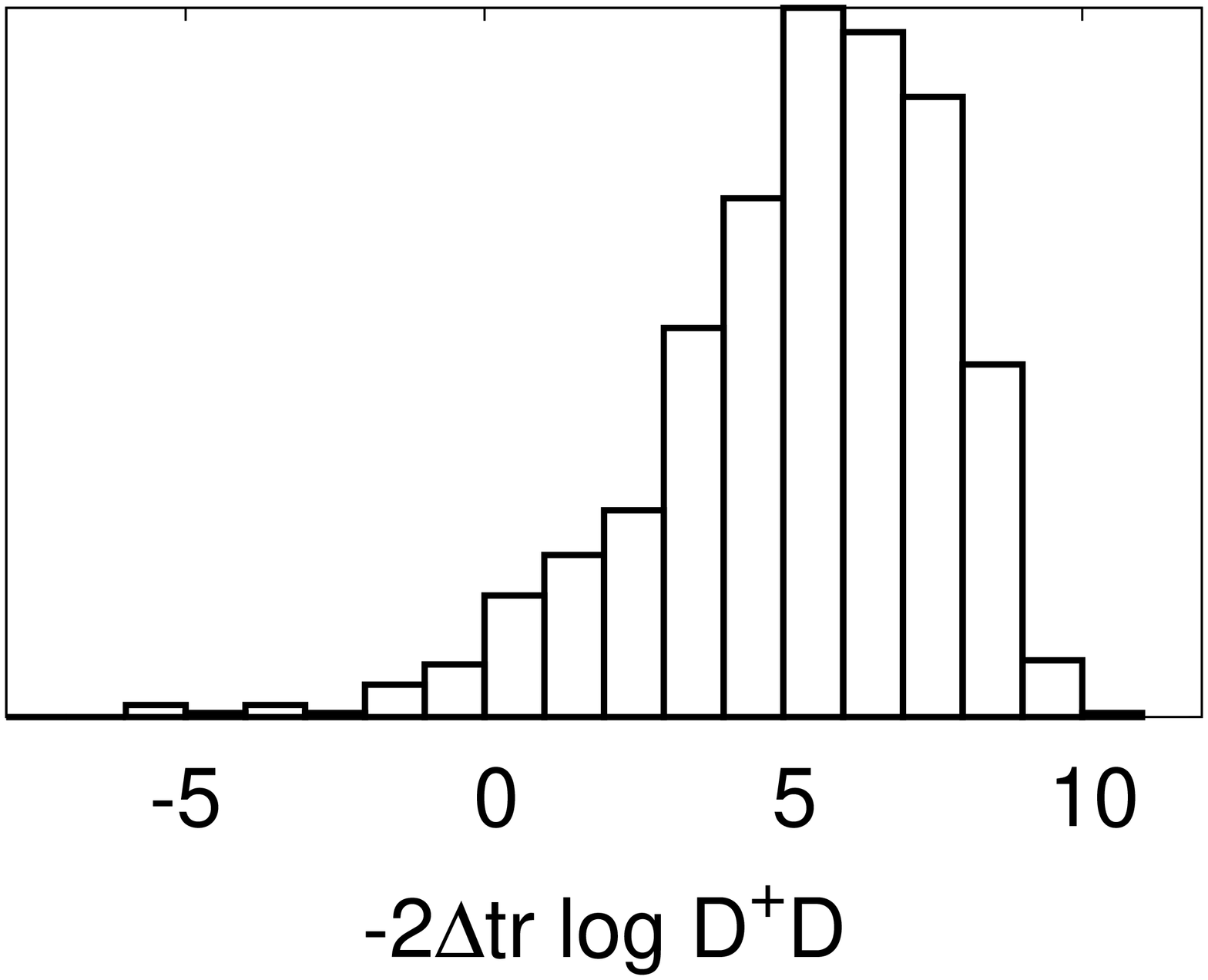}
\includegraphics[width=0.3\textwidth,angle=-90]{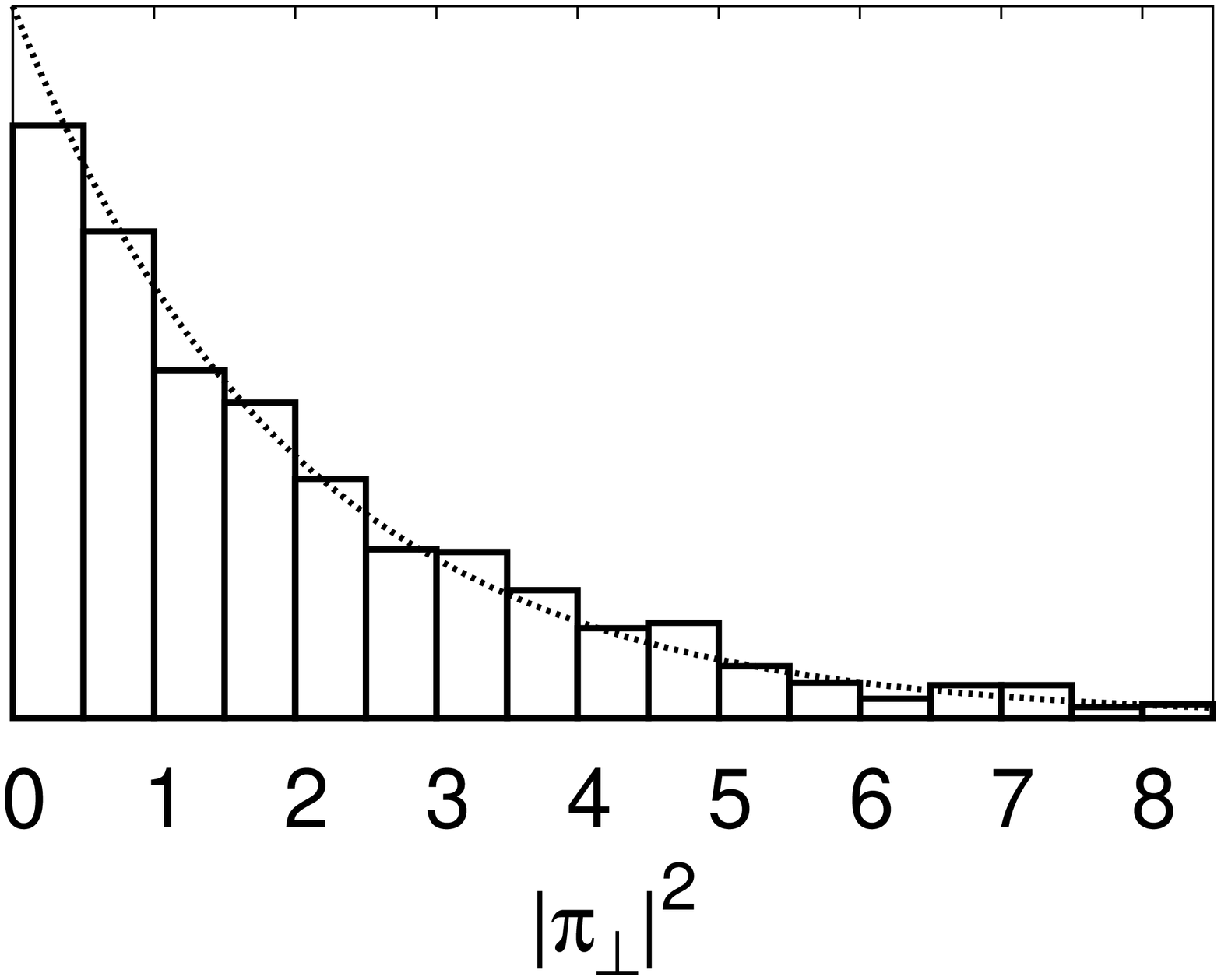}
\end{center}
\caption{The distribution of the height of the discontinuity in the 
effective action changing from the $\nu=0$ sector to $\nu=\pm1$  (upper three plots).
The lower left plot shows the change in ${\rm tr}\log D^\dagger D$, without recourse to
pseudo-fermions. The right panel contains the momentum component perpendicular to 
the surface which separates the two topological sectors. 
The topology is changed if $2\Delta S > |\pi_\perp|^2$. 
The data is from a HMC simulation on $6^4$ lattices, $a\approx0.15$fm  and $am=0.1$.
\label{fig:ds}}
\end{figure}

\begin{figure}
\begin{center}
\includegraphics[width=0.3\textwidth,angle=-90,clip]{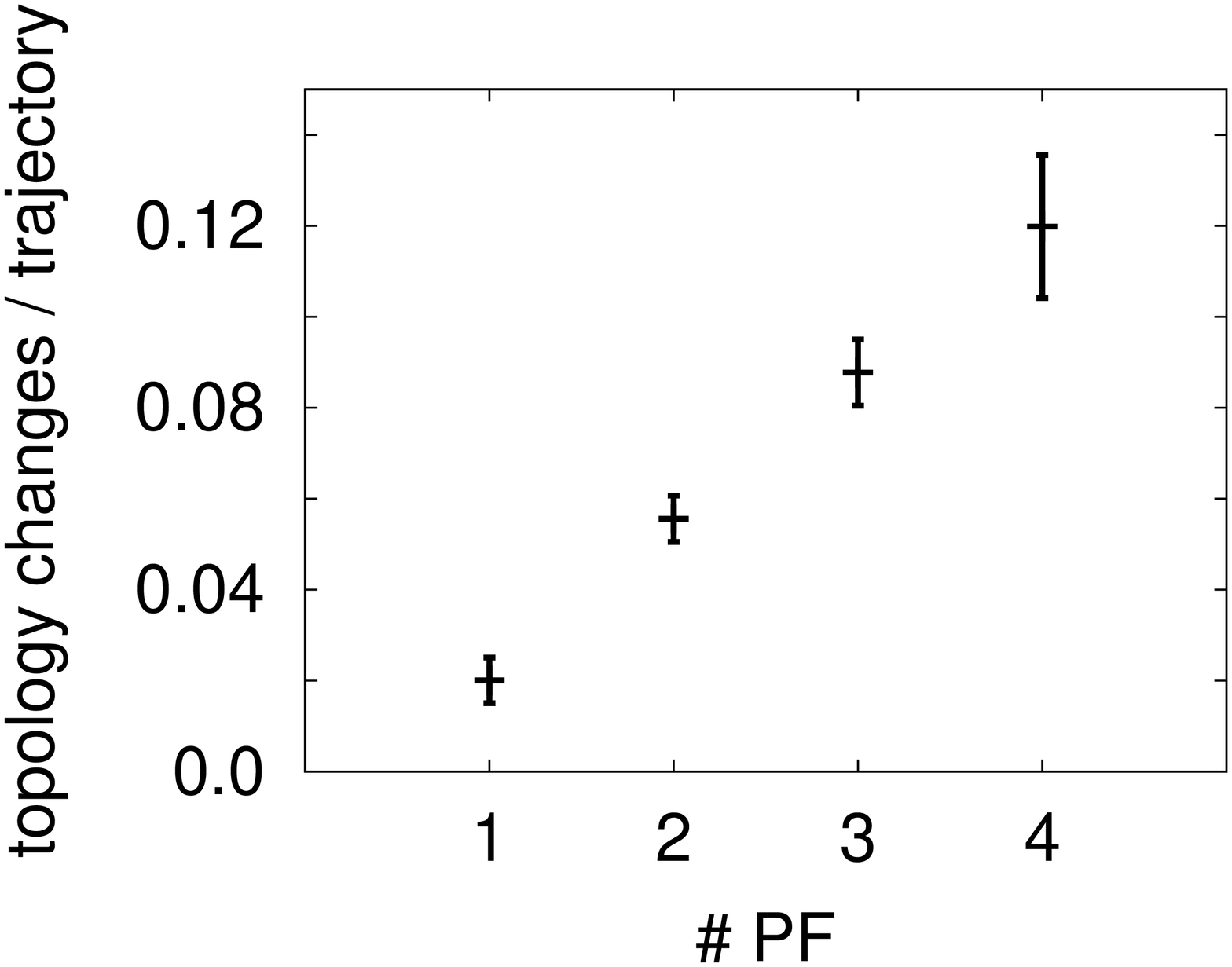}
\hspace{1cm}
\includegraphics[width=0.3\textwidth,angle=-90,clip]{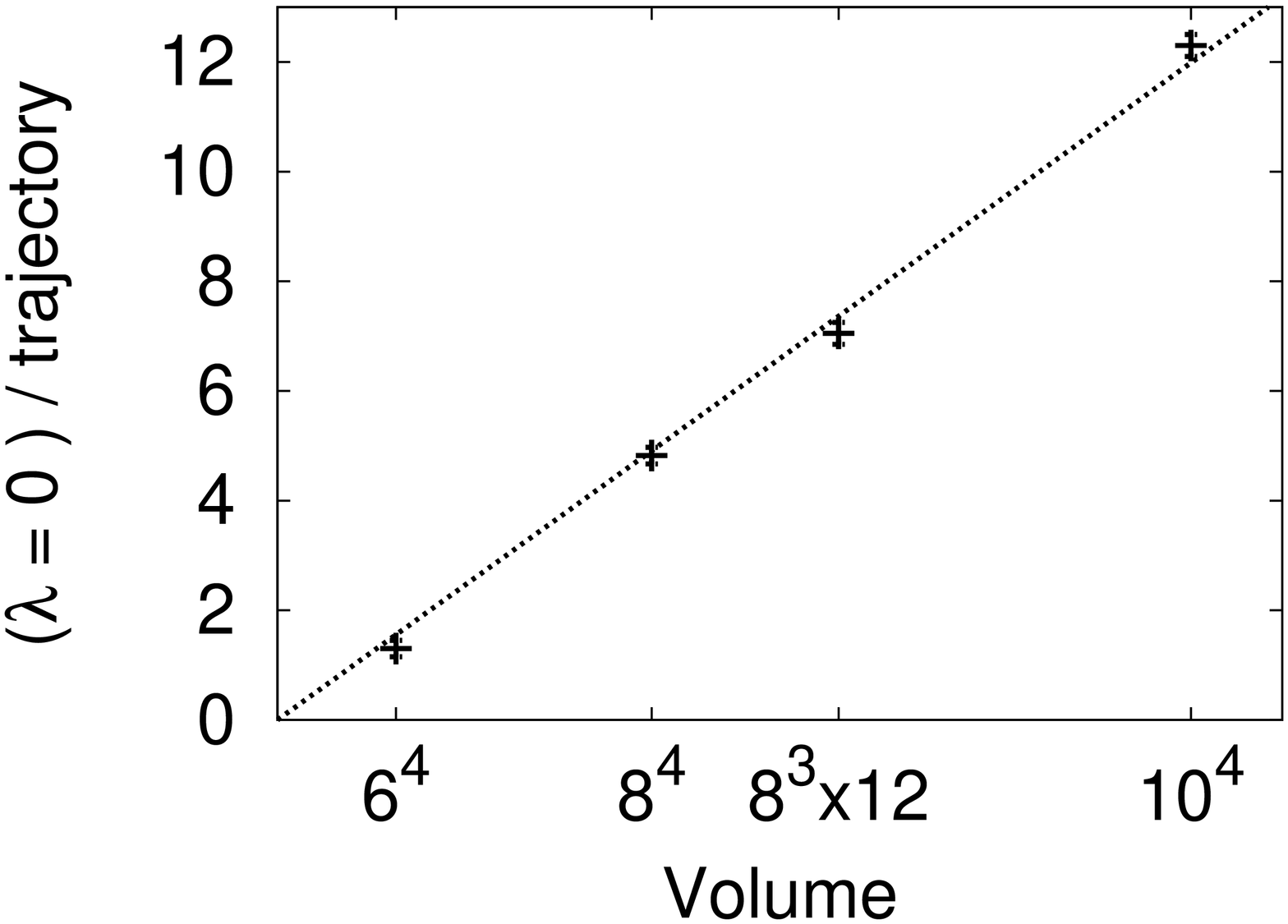}
\end{center}
\caption{\label{fig:ncross}{\bf Left}: The dependence of the tunneling rate on the number of pseudo-fermion
fields. (data set as in Fig.~{\protect \ref{fig:ds}}) {\bf Right}: Number of attempted tunnelings per trajectory
as a function of the volume. The lattice spacing in about $a\approx 0.15$fm, the quark mass $m_q\approx 100$MeV.}
\end{figure}

How does this algorithm perform at changing topological sector?
A time history of the topological charge 
from a simulation on a $8^4$ lattice at three values of the 
bare quark mass is shown in Fig.~\ref{fig:qq}. 
Whereas the charge changes frequently for the largest quark mass, the simulation 
tends to get stuck in one sector once the quark mass is lowered. 
As we will see below (Fig.~\ref{fig:ncross}) the reason is not a lack of attempts to change topology.
Rather, the discontinuity is too high to get across.
It turns out that the distribution 
of $\pi_\perp$ is not changed from the beginning to the point where the trajectory reaches
the discontinuity. As shown in Fig.~\ref{fig:ds} in the lower right panel, it still has the exponential shape
$\exp(-|\pi|^2/2)$ 
given by the initial heat bath. This makes values of  $|\pi_\perp|^2$ above 8 very unlikely independent
of the quark mass. This data, and the one used in the following discussion comes from 
a $6^4$ lattice with $a\approx0.15$~fm and $m\approx 100$~MeV. This quark mass is rather 
large and the problems described aggravate once the quark mass is lowered.

The distribution of the height of the step depends on the quark mass. We expect this because 
a smaller quark mass suppresses configurations of higher topology. However, as it 
turns out, it depends even more on how well the fermion determinant is introduced into
the functional integral.
Fig.~\ref{fig:ds}(top) shows the distribution of the height of the discontinuity 
changing from $\nu=0$ to $|\nu|=1$ for one, two
and three pseudo-fermion fields. One field yields a very wide distribution and most entries
have a value above 10 such that tunneling is virtually impossible.
Additional pseudo-fermions improve the situation; the distribution narrows. That the 
poor tunneling rate is not a problem of physics is shown in the lower left panel of 
Fig.~\ref{fig:ds}
where the discontinuity in $\tr \log D^\dagger D$ is plotted. If we could compute the 
fermion determinant  and its derivative exactly---without taking resort to pseudo-fermions---
this is the step we had to overcome. 
It does not exceed 10 such that tunneling is quite likely. The large auto-correlation
time in the topological charge is largely (if not entirely) due to the poor
approximation of the determinant by pseudo-fermions---at least as long
as the gauge action does not suppress those changes. 
Fig.~\ref{fig:ncross} demonstrates that the improved estimate indeed helps with the tunneling
rate. It shows the fraction of trajectories where the final topological charge is different
from the initial one as a function of the number of pseudo-fermion fields.
A roughly linear increase is observed.
Note that the cost of a few additional fermions is more than recouped
by the larger overall step size. That is the reason this improvement method
has received wide attention in conventional simulations with Wilson type fermions.

Smaller quark masses make the pseudo-fermion estimate of the fermion determinant even
more noisy, essentially because the conditioning number of the fermion matrix increases.
 As the previous discussion has shown, this leads to even higher steps and 
lower tunneling rates. This explains why the simulation shown in Fig.~\ref{fig:qq}---particularly for
light quarks---gets stuck for many trajectories in sectors of non-zero topology.

The dependence of the rate of change on the approximation of the determinant might
seem to interfere with detailed balance. However, detailed balance only makes a 
statement about the ratio of the probability to change from (a configuration in)
one topological sector to the other
as compared to the change in the opposite direction.
A better estimate of the determinant, however, improves the 
tunneling rate in both directions and we still get an exact algorithm.
The optimal tunneling rate is given by the exact determinant which can 
therefore serve as a guideline on the quality of the approximation.
We also note that these findings (probably) do not depend on the particular algorithm used. Any exact
algorithm has the same problems as long as the determinant is introduced just by
pseudo-fermions. However, as we have shown, different ways to introduce the 
determinant can lead to very different algorithmic performance.

A particular issue with this algorithm is the scaling with the volume.
Each time an eigenmode of the kernel operator changes sign, the height of the discontinuity
has to be computed. For that the overlap operator has to be inverted once on the crossing mode.
The cost of this inversion scales at least with the volume.
Furthermore, the number of attempted crossings turns out to be proportional 
to the volume, see Fig.~\ref{fig:ncross} on the right. This is no surprise since the number of eigenvalues is 
proportional to the volume too. The total cost of the part of the algorithm which deals
with changing topological sector therefore scales at least with $V^2$.

Whether this scaling law constitutes a problem depends on the actual number of
kernel eigenvalue crossings per trajectory at the target volume. This is mainly
a function of the density of eigenvalues near the origin which can be reduced
by various methods.
Gauge actions like the Iwasaki action or DBW2 are known to suppress the occurrence of small 
eigenmodes. This is the method of choice in the context of domain wall fermion simulations.
It turns out, however, that in particular DBW2 has a negative impact on the auto-correlation
time of the topological charge~\cite{Antonio:2005wj}.

Another strategy is to use fat links in the definition of the kernel operator. Particularly
popular are stout links~\cite{Morningstar:2003gk} because they are differentiable and
can therefore be used in molecular dynamics simulations. So far no impact of the 
smearing on the number of actual changes of topology has been observed. It is believed
that the effect of the smearing rather reduces the noise in the motion of the eigenmodes
and makes it more directed.

To summarize: The modification of the leap-frog integrator to account for the discontinuity
due to changing topological sector leads to an exact algorithm. This has been used
successfully in small volume simulations of QCD. The auto-correlation time
of the topological charge is large but can be reduced by reducing the noise in 
the estimator of the fermion determinant. The problematic point is that each attempt
to change topological sector requires one inversion of the overlap operator. The 
frequency of these attempts scales with the volume which renders this component of the
total update expensive when the lattice size is increased.

\subsection{Approximate evolution\label{sec:approx}}
In the previous section, we discussed a modification of the leap-frog in the HMC algorithm
to deal with the step in the action. The
discontinuity comes from the sign function in the definition of the overlap operator.
If instead one uses a continuous approximation to the sign function, the resulting approximate
overlap operator $D_{\rm app}$ is a continuous function of the gauge fields and
standard HMC can be used.
As long as the approximation is close enough to the exact overlap operator, it
is possible to correct for the difference between $D_{\rm app}$ and $D_{\rm ov}$ 
 at the end.

There are two different but related approaches to implement this idea~\cite{Christian:2005gd}:
The first is to rewrite the fermion determinant as
\bee
\det D_{\rm ov} = \det D_{\rm app} \det D_{\rm app}^{-1} D_{\rm ov} \ . 
\label{eq:app}
\ee
One then simulates the theory with the determinant of $ D_{\rm app}$ and 
reweights at the end to the exact overlap operator. A variant of this
approach is to 
use the determinant ratio in an additional Metropolis step at the end of each trajectory.
The second, related possibility is to perform the initial heat-bath and the 
Metropolis step at the end of the trajectory with the exact overlap operator whereas the
approximation is only used in the computation of the force during the MD evolution.

\begin{figure}
\includegraphics[width=0.3\textwidth,angle=-90]{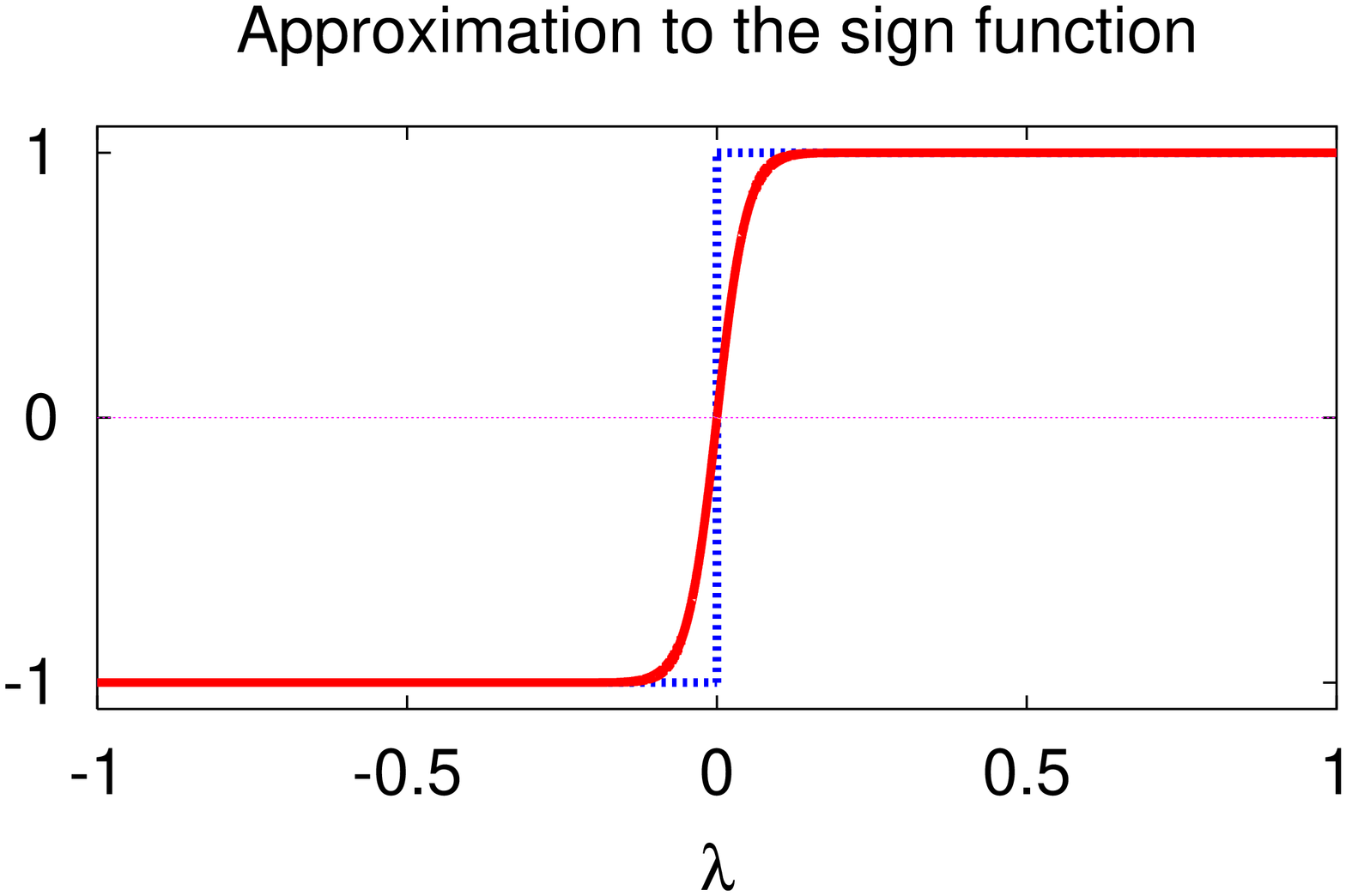}
\includegraphics[width=0.3\textwidth,angle=-90]{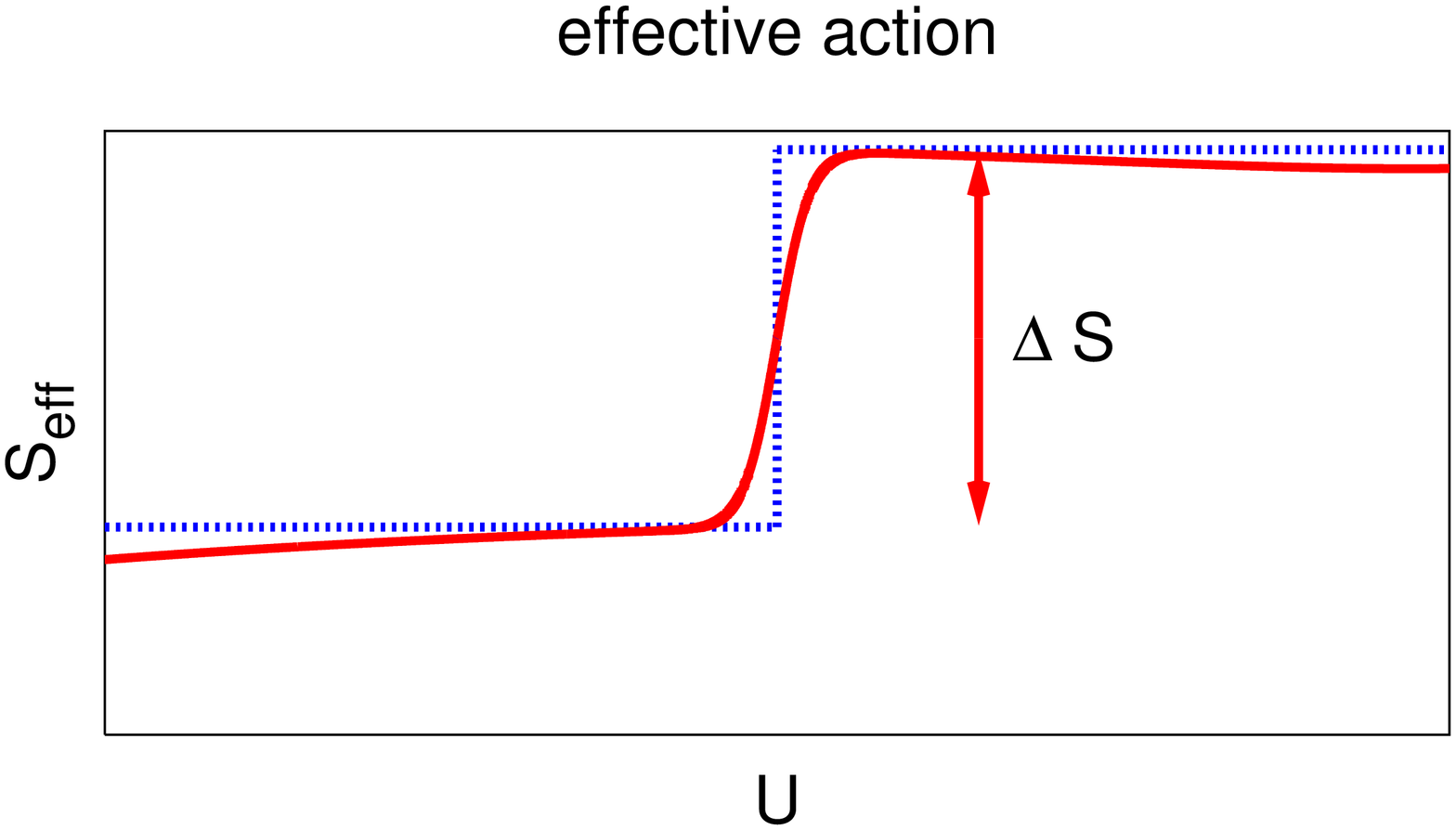}
\caption{Approximation to the sign function and the effect on the effective action.\label{fig:approx}}
\end{figure}

To understand the issues involved in this kind of simulation, let us start with the
second approach  put forward by Bode et al.~\cite{Bode:1999dd} (for a more
extensive study see \cite{Christian:2005gd,Christian:2005yp}). They use a fixed
rational approximation in the definition of the sign function during the evolution.
This leads to the situation depicted in Fig.~\ref{fig:approx} where the
Zolotarov approximation of the sign function is shown such that 
the difference for the part of the spectrum of the kernel operator with
$|\lambda|>\lambda_{\rm cut}$ is very small. 
As long as there are not too many eigenmodes with $|\lambda|<\lambda_{\rm cut}$,
the discontinuity in the overlap action is approximated by a steep section in $S_{\rm eff}$ 
which interpolates between the two levels. The width of this  section 
is roughly proportional to $\lambda_{\rm cut}$, the height proportional to $\Delta S$, the 
discontinuity in the overlap effective action. 

This poses a tuning problem. On the one hand, $D_{\rm  app}$
has to be close to the overlap operator ($\lambda_{\rm cut}$ has to be small)
to easily correct for the difference. On the other hand, a small $\lambda_{\rm cut}$ leads
to a large derivative of the effective action and to large forces in the 
molecular dynamics evolution while changing topological sector.
Therefore a small step-size is required which makes the simulation expensive.
 The analysis in the context of the FKS algorithm of Sec.~\ref{sec:FKS}
 helps here. The height of the discontinuity
and thereby the forces in the approximate action can be reduced by improving
the estimator of the fermion determinant. This leads
to smaller forces and an increased probability to accept a trajectory during
which the topology changed.

So far the method  relies on a stochastic estimate between the 
determinant of the exact and approximate overlap operator. With the specific 
choice of $D_{\rm app}$, however, the determinant ratio in Eq.~\ref{eq:app}
can be computed exactly.
It is set up such that the difference between the exact sign function
and the approximation is zero (or negligible) for $|x|>\lambda_{\rm cut}$.
From the definition of the overlap operator Eq.~\ref{eq:dov}, the following
relation is obtained
\bee
D_{\rm app} = D_{\rm ov} +
(R_0-\frac{m}{2})\sum_{i=0}^N \delta \epsilon(\lambda_i)\gamma_5 \psi_i \psi_i^\dagger \ .
\ee
with the sum running over all $N$ eigenmodes $\psi$ of the kernel operator $h$ 
with an eigenvalue of modulus smaller than $\lambda_{\rm cut}$. The 
difference between the approximate sign function and the exact one is denoted by 
$\delta \epsilon(x)$.
The ratio between the respective determinant is then given by
\bee
\det D_{\rm app} = \det  D_{\rm ov} \det_{N\times N} R \ \ \ \ \text{with} \ \  \ \
R_{ij} = 1 + (R_0-\frac{m}{2}) \delta \epsilon(\lambda_i) \psi_i^\dagger \frac{1}{\gamma_5 D_{\rm ov}} \psi_j \ .
\ee
The construction of the $N\times N$ matrix $R$ needs $N$ inversions of the overlap
operator. This method is feasible as long as $N$ is small and the fluctuations in $R$ are moderate.
Tests on $8^4$ and $10^4$ lattices are encouraging and subject of an upcoming paper~\cite{schaeferRW}.

There have been extensive tests  using approximate guidance Hamiltonians to simulate
the overlap operator in the Schwinger model.
Christian et al.~\cite{Christian:2005gd} suggested  to use a hypercubic 
Dirac operator  in the  guidance Hamiltonian. Because the hypercubic operator
is already an approximate solution of the GW equation, the overlap operator constructed
from it is similar to the kernel.
It turns out that it is not similar enough and  the authors find an acceptance rate which 
vanishes with the inverse of the volume. It becomes unacceptably small on their large $32^2$
lattices. They conclude that a better approximation to the sign function
has to be used and discuss possible forms in Ref.~\cite{Christian:2005yp}. Ref.~\cite{Volkholz:2006yh} extends this 
approach by using in the computation of the force 
a low order polynomial of the hypercubic operator which better
approximates the sign function. The volume dependence of this approach was not studied.

Overall, simulating the overlap action using an approximation seems an interesting
alternative to the modified evolution discussed in the previous section.
For differentiable approximations, standard methods apply which makes it easier to 
implement. Also the costly determination of the height of the discontinuity where
the topology changes is not necessary. However, large forces are encountered while
changing topological sectors. This requires some tuning to balance ease of simulation
(small forces) and a small difference between the two operators such that the 
reweighting is possible.
How these methods scale with the volume is difficult
to answer and needs further study. The increased density of modes due to the larger
volume probably requires  a better approximation of the overlap action which in turn
is more expensive to simulate.

\subsection{Topology conserving actions\label{sec:constopo}}
As discussed in the previous sections,
dealing with the discontinuity in the fermionic action at the interface of 
topological sectors is expensive and can lead to a $V^2$ scaling. 
An alternative approach is to fix the 
topological charge during the simulation. This eliminates the cost of determining
the height of the step in the FKS algorithm. The fixed topology can be 
an advantage, e.g. in the epsilon regime where predictions are made for 
fixed topological charge. In particular configurations with large charge are 
virtually impossible to obtain in an algorithm with tunneling because
they are suppressed by the fermion determinant. 
Unfortunately, it is unknown whether the 
individual topological sectors are connected. In the following, we will assume
that this is the case and  one can get from
any configuration of topology $Q$ to any other by continuous deformation of the 
gauge fields without leaving this particular sector. 

Still, most physical observables
are defined as averages over the full $\theta$-vacuum. 
 With no additional knowledge about the relative weights of the
topological sectors the different sets cannot sensibly be combined.
Therefore some physics, which is particularly sensitive to the topological
sampling, cannot be addressed with this method.
However, global topology is a boundary condition.
Thus for most observables,
its effect will become negligible for very large volume vanishing with $1/V$~\cite{Brower:2003yx}.
This is not as rapid as finite volume effects which  come from pions winding 
around the torus and exponentially vanish with $\exp(-m_\pi L)$.
 By comparing the 
results from simulations fixed in different topological sectors one can get a
handle on these finite volume effects. 
A recent quenched study~\cite{Galletly:2006hq}, e.g., found a rather strong dependence of the 
pseudo-scalar mass on the topological charge.

The obvious way to fix the topology is to eliminate the possibility to tunnel in the FKS algorithm.
One is still left with the reflections but the expensive determination of the height
of the discontinuity is no longer necessary. An interesting extension of this 
method~\cite{Egri:2005cx} is to compute the exact determinant ratios between
the two sides of the interface using Eq.~\ref{eq:deltadet}.
From this one can infer the relative weight of the 
two topological sectors without need of a stochastic estimate.
This again requires inversions of the overlap, but the hope is that the relative
weights can be measured more precisely than from the stochastic estimates by
pseudo-fermion fields.

Another approach which has received considerable attention is to choose an 
action such that the topological sector cannot be changed during
molecular dynamics evolution, at least with an exact integration of the equations of motion.
(The different topological sectors are separated by regions of zero weight.)
In Ref.~\cite{Hernandez:1998et} it was established that this is the case
if all plaquette variables are forced to satisfy
\bee
S_P[U]<\epsilon=\frac{1}{30}
\ee
with $S_P$ the plaquette gauge action. Later a looser bound $\epsilon\approx\frac{1}{20.5}$ could
be proved~\cite{Neuberger:1999pz}. Unfortunately, it turned out that 
the simulation of these gauge actions is difficult~\cite{Bietenholz:2005rd,Fukaya:2005cw} in
particular at relevant lattice spacings. 

In a similar spirit is a method suggested by Vranas in the context of domain wall fermions~\cite{Vranas:1999rz}.
The idea is to introduce the determinant of the kernel operator $\det(h(-R_0))$ into the 
functional integral. This determinant is zero where one of the eigenmodes of $h(-R_0)$ 
changes sign, i.e. where the topology as seen by the overlap operator
changes.  The determinant therefore introduces a repulsive potential against
the change of the topological sector and eliminates these changes completely if one
integrates the equations of motion precisely enough.
The effect of $\det(h(-R_0))$ is expected to vanish in the continuum limit since it
corresponds to a fermion at large negative mass at the cut-off scale, essentially
acting like a modification of the gauge action.

JLQCD has started to use a variant of this trick in their large scale simulations
with dynamical overlap fermions. They use the following ratio of determinants to 
prevent the trajectories from changing topology
\bee
\frac{\det h^2(-R_0)}{\det (h^2(-R_0) +\mu^2)}
\ee
with tunable (twisted) mass parameter $\mu$. This term still is zero only where
$h(-R_0)$ has a zero mode, however, the effect of the higher modes largely cancels.
This cancellation is complete if $\mu=0$ but then the effect which suppresses
the changing of topology vanishes too. The advantage of fixing the topology in this way is that it proved feasible to simulate (without
the overlap determinant) in the targeted  range of lattice spacings and volumes~\cite{Fukaya:2006vs}.

At this conference, first results with this method were presented. The simulations
were done on $16^3\times 32$ lattices at a lattice spacing around 0.10fm with
various values for the sea quark mass~\cite{JLQCD}.

\section{Summary}
For chiral fermions, the index theorem already holds at finite lattice spacing. This
causes the fermionic action to be discontinuous where the topological charge defined
by the index changes. The preferred way to deal with this situation depends on 
the importance of the topological charge in the observable under investigation. Three
distinct approaches have been discussed, all based on the HMC algorithm.

The most radical solution is the one discussed last. The action is modified in such
a way that a continuous modification of the gauge fields does not lead to a change
in the topological charge. This is achieved by introducing the 
determinant of the kernel of the overlap operator. 
Therefore no discontinuity is encountered and standard
methods can be used to simulate the theory. 
For most observables,  the effect of  fixed topology is expected to vanish 
like $1/V$ for $V\to \infty$.
How large a volume is needed for this to be under control 
will have to be investigated in explicit simulations.  
Such simulations are currently under way and the JLQCD collaboration has
already presented first results at this conference.

The other two approaches to dynamical overlap fermions allow for topology to change.
The first is to modify the molecular dynamics evolution and thereby integrate
the discontinuity exactly. Each time an eigenmode of the kernel operator 
changes sign, the height of the step has to be determined which costs one 
inversion of the overlap operator. This leads to $V^2$ scaling of that part of the 
algorithm because its cost scales with the volume and so does its frequency.
The advantage is that no large forces occur because of changing topological 
sector. Also the topological charge history is very well under control.

An alternative approach is to approximate the discontinuity and then correct for
this either by reweighting or a Metropolis step. This works if one can find
an approximate action which on the one hand is close enough to the overlap action
for the correction term to be small.
On the other hand, the forces encountered while changing topological sectors have
to be small enough to be able to integrate the equations of motion to good accuracy with a reasonable
step size. This poses a tuning problem and again simulations have to show how well
this method works in practice. Studies in the Schwinger model are encouraging.

The history of the topological charge is of particular interest in dynamical simulations
of chiral fermions. We were able to demonstrate that the tunneling rate depends crucially on
the way the fermion determinant is introduced into the functional integral. Large
noise in its estimator leads to long auto-correlation times in the topology. So far,
only  Hasenbusch's mass preconditioning was studied and found to be
very effective. Other methods might render even better results.

Dynamical overlap simulations are still very expensive. However, computer power will
grow and there certainly will be better algorithms too. The benefit of these
efforts is a theoretically clean description of QCD with chiral symmetry at
finite lattice spacing.

\acknowledgments
I want to thank T.~DeGrand and K.~Jansen for many conversations and their comments on 
the manuscript.
Discussion and correspondence with A.~Borici, N.~Cundy, T.~Draper, S.~Hashimoto,
 C.~Lang, Y.~Shamir is also gratefully acknowledged.

\bibliographystyle{JHEP}
\bibliography{www}

\providecommand{\href}[2]{#2}\begingroup\raggedright\begin{thebibliography}{10}

\bibitem{Luscher:1998pq}
M.~L{\"u}scher, {\em Phys. Lett.} {\bf B428} (1998) 342--345,
  [\href{http://xxx.lanl.gov/abs/hep-lat/9802011}{{\tt hep-lat/9802011}}].

\bibitem{Ginsparg:1981bj}
P.~H. Ginsparg and K.~G. Wilson, {\em Phys. Rev.} {\bf D25} (1982) 2649.

\bibitem{Hasenfratz:1998ri}
P.~Hasenfratz, V.~Laliena, and F.~Niedermayer, {\em Phys. Lett.} {\bf B427}
  (1998) 125--131, [\href{http://xxx.lanl.gov/abs/hep-lat/9801021}{{\tt
  hep-lat/9801021}}].

\bibitem{Hasenfratz:1998jp}
P.~Hasenfratz, {\em Nucl. Phys.} {\bf B525} (1998) 401--409,
  [\href{http://xxx.lanl.gov/abs/hep-lat/9802007}{{\tt hep-lat/9802007}}].

\bibitem{Kaplan:1992bt}
D.~B. Kaplan, {\em Phys. Lett.} {\bf B288} (1992) 342--347,
  [\href{http://xxx.lanl.gov/abs/hep-lat/9206013}{{\tt hep-lat/9206013}}].

\bibitem{Shamir:1993zy}
Y.~Shamir, {\em Nucl. Phys.} {\bf B406} (1993) 90--106,
  [\href{http://xxx.lanl.gov/abs/hep-lat/9303005}{{\tt hep-lat/9303005}}].

\bibitem{Neuberger:1997fp}
H.~Neuberger, {\em Phys. Lett.} {\bf B417} (1998) 141--144,
  [\href{http://xxx.lanl.gov/abs/hep-lat/9707022}{{\tt hep-lat/9707022}}].

\bibitem{Hasenfratz:1993sp}
P.~Hasenfratz and F.~Niedermayer, {\em Nucl. Phys.} {\bf B414} (1994) 785--814,
  [\href{http://xxx.lanl.gov/abs/hep-lat/9308004}{{\tt hep-lat/9308004}}].

\bibitem{Hasenfratz:2000xz}
P.~Hasenfratz, {\em et~al.} {\em Int. J. Mod. Phys.} {\bf C12} (2001) 691--708,
  [\href{http://xxx.lanl.gov/abs/hep-lat/0003013}{{\tt hep-lat/0003013}}].

\bibitem{ClarkPlenary}
M.~Clark, {\em The rational Hybrid Monte Carlo algorithm}, \pos{PoS(LAT2006)}.

\bibitem{GiustiPlenary}
L.~Giusti, {\em Light dynamical fermions on the lattice: toward the chiral regime of QCD}, \pos{PoS(LAT2006)}.

\bibitem{Hasenfratz:2005tt}
A.~Hasenfratz, P.~Hasenfratz, and F.~Niedermayer, {\em Phys. Rev.} {\bf D72}
  (2005) 114508, [\href{http://xxx.lanl.gov/abs/hep-lat/0506024}{{\tt
  hep-lat/0506024}}].

\bibitem{Lang:2005jz}
C.~B. Lang, P.~Majumdar, and W.~Ortner, {\em Phys. Rev.} {\bf D73} (2006)
  034507, [\href{http://xxx.lanl.gov/abs/hep-lat/0512014}{{\tt
  hep-lat/0512014}}].

\bibitem{Hernandez:1998et}
P.~Hernandez, K.~Jansen, and M.~L{\"u}scher, {\em Nucl. Phys.} {\bf B552} (1999)
  363--378, [\href{http://xxx.lanl.gov/abs/hep-lat/9808010}{{\tt
  hep-lat/9808010}}].

\bibitem{DeGrand:2005vb}
T.~A. DeGrand and S.~Schaefer, {\em Phys. Rev.} {\bf D72} (2005) 054503,
  [\href{http://xxx.lanl.gov/abs/hep-lat/0506021}{{\tt hep-lat/0506021}}].

\bibitem{Chiarappa:2006hz}
T.~Chiarappa, {\em et~al.} \href{http://xxx.lanl.gov/abs/hep-lat/0609023}{{\tt
  hep-lat/0609023}}.

\bibitem{Duane:1987de}
S.~Duane, A.~D. Kennedy, B.~J. Pendleton, and D.~Roweth, {\em Phys. Lett.} {\bf
  B195} (1987) 216--222.

\bibitem{Gottlieb:1987mq}
S.~A. Gottlieb, W.~Liu, D.~Toussaint, R.~L. Renken, and R.~L. Sugar, {\em Phys.
  Rev.} {\bf D35} (1987) 2531--2542.

\bibitem{Fodor:2003bh}
Z.~Fodor, S.~D. Katz, and K.~K. Szabo, {\em JHEP} {\bf 08} (2004) 003,
  [\href{http://xxx.lanl.gov/abs/hep-lat/0311010}{{\tt hep-lat/0311010}}].

\bibitem{Fodor:2004wx}
Z.~Fodor, S.~D. Katz, and K.~K. Szabo, {\em Nucl. Phys. Proc. Suppl.} {\bf 140}
  (2005) 704--706, [\href{http://xxx.lanl.gov/abs/hep-lat/0409070}{{\tt
  hep-lat/0409070}}].

\bibitem{Hasenfratz:2002ym}
A.~Hasenfratz and A.~Alexandru, {\em Phys. Rev.} {\bf D65} (2002) 114506,
  [\href{http://xxx.lanl.gov/abs/hep-lat/0203026}{{\tt hep-lat/0203026}}].

\bibitem{Hasenbusch:2001ne}
M.~Hasenbusch, {\em Phys. Lett.} {\bf B519} (2001) 177--182,
  [\href{http://xxx.lanl.gov/abs/hep-lat/0107019}{{\tt hep-lat/0107019}}].

\bibitem{Hasenbusch:2002ai}
M.~Hasenbusch and K.~Jansen, {\em Nucl. Phys.} {\bf B659} (2003) 299--320,
  [\href{http://xxx.lanl.gov/abs/hep-lat/0211042}{{\tt hep-lat/0211042}}].

\bibitem{Hasenbusch:1998yb}
M.~Hasenbusch, {\em Phys. Rev.} {\bf D59} (1999) 054505,
  [\href{http://xxx.lanl.gov/abs/hep-lat/9807031}{{\tt hep-lat/9807031}}].

\bibitem{Clark:2006fx}
M.~A. Clark and A.~D. Kennedy,
  \href{http://xxx.lanl.gov/abs/hep-lat/0608015}{{\tt hep-lat/0608015}}.

\bibitem{DeGrand:2004nq}
T.~A. DeGrand and S.~Schaefer, {\em Phys. Rev.} {\bf D71} (2005) 034507,
  [\href{http://xxx.lanl.gov/abs/hep-lat/0412005}{{\tt hep-lat/0412005}}].

  
\bibitem{Cundy:2004xf}
N.~Cundy, S.~Krieg, A.~Frommer, T.~Lippert, and K.~Schilling, {\em Nucl. Phys.
  Proc. Suppl.} {\bf 140} (2005) 841--843,
  [\href{http://xxx.lanl.gov/abs/hep-lat/0409029}{{\tt hep-lat/0409029}}].

\bibitem{Cundy:2005pi}
N.~Cundy, {\em et~al.} \href{http://xxx.lanl.gov/abs/hep-lat/0502007}{{\tt
  hep-lat/0502007}}.

\bibitem{Schaefer:2005qg}
S.~Schaefer and T.~A. DeGrand, {\em PoS} {\bf LAT2005} (2006) 140,
  [\href{http://xxx.lanl.gov/abs/hep-lat/0508025}{{\tt hep-lat/0508025}}].

\bibitem{Cundy:2005mn}
N.~Cundy, S.~Krieg, and T.~Lippert, {\em PoS} {\bf LAT2005} (2006) 107,
  [\href{http://xxx.lanl.gov/abs/hep-lat/0511044}{{\tt hep-lat/0511044}}].

\bibitem{DeGrand:2006ws}
T.~DeGrand and S.~Schaefer, {\em JHEP} {\bf 0607} (2006)
  020, [\href{http://xxx.lanl.gov/abs/hep-lat/0604015}{{\tt hep-lat/0604015}}].

\bibitem{DeGrand:2006uy}
T.~DeGrand, R.~Hoffmann, S.~Schaefer, and Z.~Liu,
  \href{http://xxx.lanl.gov/abs/hep-th/0605147}{{\tt hep-th/0605147}}.

\bibitem{DeGrand:2006nv}
T.~DeGrand, Z.~Liu, and S.~Schaefer,
  \href{http://xxx.lanl.gov/abs/hep-lat/0608019}{{\tt hep-lat/0608019}}.

\bibitem{Antonio:2005wj}
{\bf RBC and UKQCD} Collaboration, D.~J. Antonio, {\em et~al.} {\em PoS} {\bf
  LAT2005} (2006) 135.

\bibitem{Morningstar:2003gk}
C.~Morningstar and M.~J. Peardon, {\em Phys. Rev.} {\bf D69} (2004) 054501,
  [\href{http://xxx.lanl.gov/abs/hep-lat/0311018}{{\tt hep-lat/0311018}}].

\bibitem{Christian:2005gd}
N.~Christian, K.~Jansen, K.-I. Nagai, and B.~Pollakowski, {\em PoS} {\bf
  LAT2005} (2006) 239, [\href{http://xxx.lanl.gov/abs/hep-lat/0509174}{{\tt
  hep-lat/0509174}}].

\bibitem{Bode:1999dd}
A.~Bode, U.~M. Heller, R.~G. Edwards, and R.~Narayanan,
  \href{http://xxx.lanl.gov/abs/hep-lat/9912043}{{\tt hep-lat/9912043}}.

\bibitem{Christian:2005yp}
N.~Christian, K.~Jansen, K.~Nagai, and B.~Pollakowski, {\em Nucl. Phys.} {\bf
  B739} (2006) 60--84, [\href{http://xxx.lanl.gov/abs/hep-lat/0510047}{{\tt
  hep-lat/0510047}}].

\bibitem{schaeferRW}
S.~Schaefer, {\em in preparation}.

\bibitem{Volkholz:2006yh}
J.~Volkholz, W.~Bietenholz, and S.~Shcheredin,
  \href{http://xxx.lanl.gov/abs/hep-lat/0609003}{{\tt hep-lat/0609003}}.

\bibitem{Brower:2003yx}
R.~Brower, S.~Chandrasekharan, J.~W. Negele, and U.~J. Wiese, {\em Phys. Lett.}
  {\bf B560} (2003) 64--74,
  [\href{http://xxx.lanl.gov/abs/hep-lat/0302005}{{\tt hep-lat/0302005}}].

\bibitem{Galletly:2006hq}
D.~Galletly, {\em et~al.} \href{http://xxx.lanl.gov/abs/hep-lat/0607024}{{\tt
  hep-lat/0607024}}.

\bibitem{Egri:2005cx}
G.~I. Egri, Z.~Fodor, S.~D. Katz, and K.~K. Szabo, {\em JHEP} {\bf 01} (2006)
  049, [\href{http://xxx.lanl.gov/abs/hep-lat/0510117}{{\tt hep-lat/0510117}}].

\bibitem{Neuberger:1999pz}
H.~Neuberger, {\em Phys. Rev.} {\bf D61} (2000) 085015,
  [\href{http://xxx.lanl.gov/abs/hep-lat/9911004}{{\tt hep-lat/9911004}}].

\bibitem{Bietenholz:2005rd}
W.~Bietenholz, {\em et~al.} {\em JHEP} {\bf 03} (2006) 017,
  [\href{http://xxx.lanl.gov/abs/hep-lat/0511016}{{\tt hep-lat/0511016}}].

\bibitem{Fukaya:2005cw}
H.~Fukaya, S.~Hashimoto, T.~Hirohashi, K.~Ogawa, and T.~Onogi, {\em Phys. Rev.}
  {\bf D73} (2006) 014503, [\href{http://xxx.lanl.gov/abs/hep-lat/0510116}{{\tt
  hep-lat/0510116}}].

\bibitem{Vranas:1999rz}
P.~M. Vranas, \href{http://xxx.lanl.gov/abs/hep-lat/0001006}{{\tt
  hep-lat/0001006}}.

\bibitem{Fukaya:2006vs}
{\bf JLQCD} Collaboration, H.~Fukaya {\em et~al.},
  \href{http://xxx.lanl.gov/abs/hep-lat/0607020}{{\tt hep-lat/0607020}}.

\bibitem{JLQCD}
{\bf JLQCD} Collaboration, {\em contributions to these proceedings by   H.~Fukaya, S.~Hashimoto, T.~Kaneko, H.~Matsufuru, M.~Okamoto, N.~Yamada} , \pos{PoS(LAT2006)}.

\end{thebibliography}\endgroup

\end{document}